\documentclass[12pt]{article}
\usepackage{amsmath}
\usepackage{mathtools}
\usepackage{graphicx}
\usepackage{amssymb}
\usepackage{graphicx}
\usepackage{epstopdf}
\usepackage{geometry}
\usepackage{inputenc}
\usepackage{amssymb}
\usepackage{wasysym}
\usepackage{latexsym}
\usepackage{appendix}
\usepackage{array}
\usepackage{indentfirst}
\usepackage{multicol}
\usepackage[T1]{fontenc} 
\graphicspath{{Graphs/}}
\usepackage{geometry}
\begin{document}

\date{}
\title{Disturbing the Peace: Anatomy of the Hostile Takeover of China Vanke Co.}
\author{Taurai Muvunza\footnote{Tsinghua-Berkeley Shenzhen Institute. Tsinghua Shenzhen International Graduate School. Shenzhen, 518055, China. Email: gnf18@mails.tsinghua.edu.cn. The author wishes to thank an 'anonymous' employee from Vanke for providing helpful comments}\\
\and Terrill Frantz\footnote{Professor of eBusiness and Cybersecurity, Harrisburg University of Science and Technology, Pennsylvania; U.S. Email: terrill@org-sim.com. }\\}
\maketitle
\begin{abstract}
Wang Shi, a business mogul who created his empire of wealth from scratch, relished in his fame and basked in the glory of his affluent business. Nothing lasts forever! After mastering the turbulent business of real estate development in his country and therefore enjoying a rising and robust stock price, China Vanke Co. Ltd (“Vanke”) founder and Chairman of the Board of Directors, Wang Shi was suddenly presented with a scathing notice from the Hong Kong Stock Exchange: rival Baoneng Group (“Baoneng”) filed the regulatory documentation indicating that it had nicodemously acquired 5\% of his company and was looking to buy more. Vanke case became brutal and sparked national controversy over corporate governance and the role of Chinese government in capital markets.\\

\textbf{Keywords:} Vanke; Baoneng; corporate governance; hostile takeover; China
\end{abstract}

 \newpage
\begin{multicols}{2}
 
\section{Introduction}
Baoneng is a fierce rival that does business in a way Wang calls “vulgar”; in no way would he allow his company and its stellar reputation and glory be plundered by an ensuing majority ownership and thus control over Vanke by Baoneng’s predatory founder and 46 year old CEO Yao Zhenhua.  The tranquility of business at Vanke and the harmony in China’s nascent world of M\&A was rattled in their core. Even the Chinese government disapprovingly took notice. Wang Shi labelled the takeover not only hostile but also ‘predatory’ and further denounced an attempt by Baoneng to become the major shareholder stating that Shenzhen Jushenghua Industrial Development Co., ("Jushenghua"), Baoneng’s subsidiary, used reckless debt to finance the share acquisition. This marked the brewing of a takeover battle, which gained attention in mainstream media, boardrooms and further attracted a consortium of companies to the battleground for control.\\

\section{History of Vanke}

Wang Shi founded China Vanke, which boasts over thirty years of experience in the housing market, in 1988 in Shenzhen. The 1982 draft of the sixth Five Year Plan under the leadership of Deng Xiaoping focused on reform and premised on progress towards a market economy, at the same time opening a window for the then 1988 provisional regulations on private enterprises that allowed the legitimization of sole proprietorship, partnerships and limited liability corporations. The same wave of reform saw the inception of conglomerates that have now become leading brands in today’s Chinese markets. When Vanke was established, real estate and housing in China had not gained traction and average income per capita was below 2 000 yuan. Low demand for housing is attributed to this phenomenon where China had a large rural population accounting for over 80\% in early 1980s and roughly, 20\% lived in the cities. This  curtailed growth in real estate sector. Vanke, a first mover in real estate, consolidated market share and as GDP grew exponentially each year, the demand for houSsing and real estate development boomed.\\
Vanke’s mission has three focus areas. Firstly, the company has defined itself as a Global Residential Developer. A large portion of yearly sales are driven by this category through providing U8 decoration system, BIM based quality control and refabricated housing. Secondly, Vanke markets itself as an integrated urban property service provider which includes shopping malls, integrated complex development, Vanke Education, Vanke Resort, and long term leasing apartments. Lastly, Vanke Worldwide is also one of the oldest focus areas of the company. Having established operations in more than 71 cities in China, Vanke Worldwide Strategy has enabled the company to cover five cities outside China including Singapore, San Francisco, New York and London. Real estate started experiencing significant growth around 1990 largely because of opening up and reform agenda. After Shanghai and Shenzhen stock exchanges were set up in 1990 and 1991, respectively, Vanke was one of the first companies to be listed in 1991 on Shenzhen Stock Exchange. In 1993, Vanke also issued B shares and established residential property development as its core business. Revenue and profit growth for Vanke averaged 29.1\% and 31.3\% compounded yearly from 1991, and the shares increased value correspondingly. From 1990, there was a steady growth in urban population. Between 1990 and 2000, urban population surged from 28\% to 36\% and by 2007; urban population was approaching 50\%. This increased the demand for housing as more and more people relocated to urban areas in search for jobs and a better life. Consequently, middle-income households grew faster and the number of households who could afford to own a house increased as wages soured. Urbanization became the driver for real estate growth. In addition to that, the government relaxed regulations on land, for example, Article 64 of the Property Rights Law made it possible to own estates and immovable property, which opened doors for real estate companies to borrow excessively from banks and buy land from government for development. Total revenue hit 100 billion yuan in 2010 and Vanke has continued to enjoy enviable profits.  In 2015, Vanke posted 261.47 billion yuan in revenue (see Figure 1), a year on year growth of 14.3\% making it an industry leader in China based on sales revenue. In July 2016, the company was listed on the “Fortune Global 500” by Fortune Magazine for the first time, ranked 356th. \\
Other Chinese developers that appeared on the list include Wanda and Evergrande. Vanke’s success cannot only be attributed to China’s rising middle income and urbanization, but also to Wang Shi’s strategy to enter new international territories such as Singapore, United States and United Kingdom. China Vanke entered the U.S. when the real estate market was recovering from the 2008 recession and its investments have been made during a period with reasonable prices. Major Chinese developers such as Wanda and Greenland Group have targeted Chinese immigrants in New York and California as their customer base. They have also invested in office property acquisitions, schools and hotels. According to Wang Shi, entering US market gives Vanke an opportunity to learn and understand business models in a developed market and gain management experience through project cooperation with other US developers. 
In 2013, Vanke partnered with Tishman Speyer and another US developer, Hines, to develop a San Francisco condominium project which became one of the early moves of Vanke in U.S. In September 2016, in a bid to step up its overseas investments from 5\% to the 20\% target, the company agreed to purchase Ryder Court in Mayfair, an office building located in Central London for GBP115 million. The property was acquired from UK-based Henderson Global Investors whose shareholders had mounting uncertainties in the real estate market after Brexit vote. Despite speculation of a decline in housing market due to uncertainties following Brexit, Vanke's overseas investments have paid off. It improved the company’s performance, positioned it to be the market leader in China’s real estate - making it one of the attractive companies on the stock market. 
Individual investors and corporations began to regard Vanke as an investment choice after the company outperformed all other developers consecutively in China. However, when China's economy reverted to a "New Normal", a dilemna occured on the markets: stocks were overvalued and economic performance could not match the rising stock prices. The government encouraged companies to prop up the stock market. New regulation that was passed to allow pension funds and insurance companies to invest in bonds and money market instruments provided a leeway for real estate companies to access capital easily. Pension funds and insurance companies which had huge sums of cash reserves also snapped the opportunity to venture in trading high risk assets. On the market, amongst real estate companies, Vanke stood out due to its stunning performance and this triggered a takeover attempt.\\
\end{multicols}
\begin{figure}[h]
\caption{Vanke's Annual Sales Revenue}
\includegraphics [scale=1]{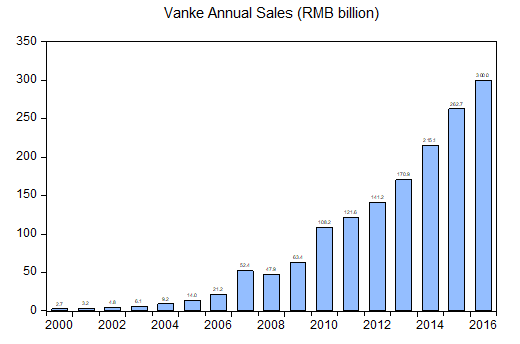} 
\end{figure}
\begin{multicols}{2}
\section{The Battle for Control 1: Baoneng vs Vanke}
Vanke had not encountered a takeover attempt and battle for control since its founding. The only time there was a tussle was on March 30, 1994 when Junan Securities representing 4 big investors purchased large volume of stocks of Vanke on the secondary market. The stock market in China was barely 4 years old and the regulations vague. Wang Shi considered it as a potential takeover. Vanke shares were suspended and Wang Shi met with the investors from Jinan Securities. On April 4, Wang Shi announced victory and the shares resumed trading. Despite the potential takeover, Wang Shi did not consider consolidating his ownership, which he gave up when the company went public in 1991, instead, shares kept floating on the market – posing a threat for another takeover. In 2000, Vanke welcomed China Resources (Holding) Co. Ltd ("China Resources"), (see Figure 2) as a major shareholder. China Resources controlled a stake of 14.89\%.
On July 25, 2016, Wang Shi woke up to discover that 15.04\% of Vanke's A shares had been purchased by Foresea Life Insurance Ltd ("Foresea Life"), at an average of 14.4 yuan per piece since July 10th. At that time, Foresea Life was 20\% owned by Jushenghua which in turn was 99\% owned by Baoneng Group (see Figure 2). Wang Shi called the purchase an attempt for hostile takeover and rejected Jushenghua and Foresea Life as major shareholders of Vanke, citing that the share purchase was bought by short-term debt, which could further negatively affect Vanke’s capital structure. Vanke held an urgent meeting to plan against the takeover and battle Baoneng. China Resources had been Vanke’s largest shareholder for 15 years, would the state-owned company rescue Vanke from the battle? \\ 
\end{multicols}
\begin{center}
\begin{figure}[h]
\caption{Ownership Structure of China Resources and Baoneng}
\includegraphics [scale=0.45]{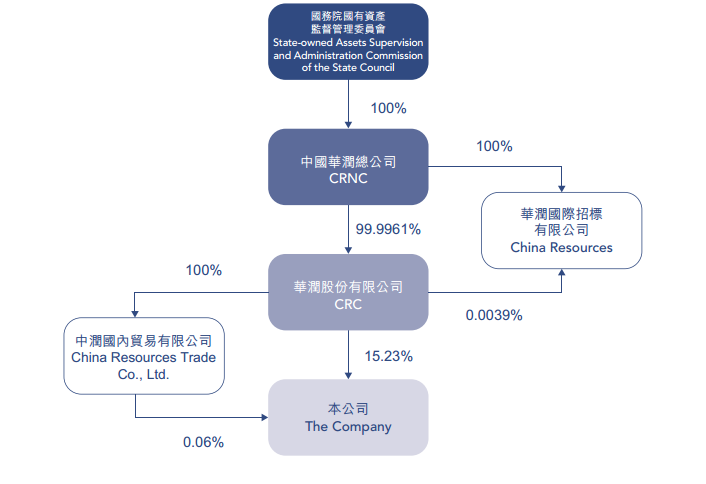}  \includegraphics[scale=0.45]{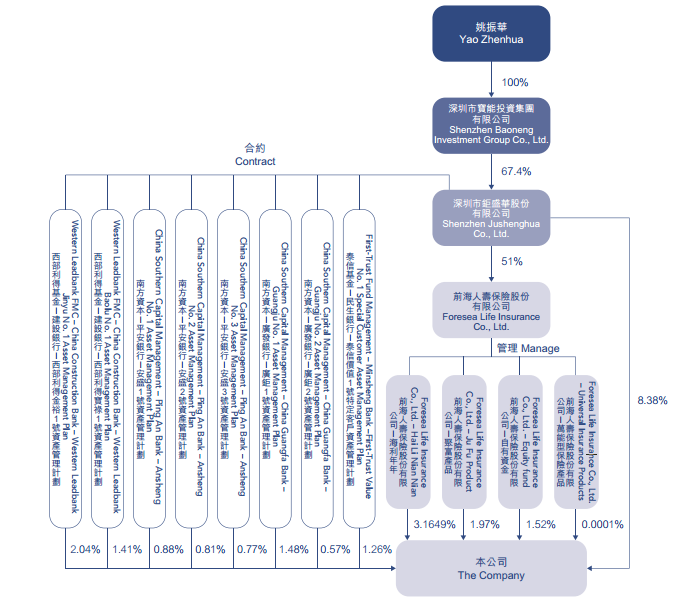}
\end{figure}
\end{center}
\begin{multicols}{2}
On September 1, China Resources increased controlling stake to 15.29\% at a cost of 500 million yuan. Baoneng Group continued to use its subsidiaries to purchase shares and consolidate its position as the largest shareholder of Vanke.   Vanke's stock price shot up rythmically in response to the takeover and within two weeks of December, the stock rose 33\% to 22.21 yuan a piece, increasing the market capitalization of the developer to a whooping 245.5 billion yuan. Meanwhile, An Bang Insurance Group, a global insurance company with total assets of nearly 1971 billion yuan was Vanke’s second largest shareholder. It raised its stake from 4.4\% to 5\% in December after Jushenghua and Foresea Life had lifted their stake twice on December 4 and 10th from 15.29\% to 22.25\% at a total average cost of 7.46 billion yuan. Total number of shares owned by Foresea Life and Jushenghua stood at 2.21 billion, acquired at a staggering cost of 35 billion yuan (Security Times). Baoneng stated the controversial acquisition was motivated by Vanke’s performance and further expressed that the continued investment in Vanke was to hid a call by government to prop up stock markets. As the battle for control continued through impulsive purchases of Vanke’s stocks, China Securities Regulatory Commission ("CSRC"); the regulatory arm of government for securities,  reported they would not intervene as long as the players abide by rules. Vanke would have welcomed the intervention of government to cool down the heat on the stock market. Although investors were taking the risk to purchase Vanke’s expensive shares which were trading at 3 times book value, they also feared that the stocks were overvalued and did not reflect the actual performance of the company. Furthermore, uncertainty over Vanke’s reaction to reverse the takeover increased. The effect would send the prices plunging down significantly. International watchdogs for investors began to offer warnings about the future of Vanke. For example, on December 16, Morgan Stanley swapped Vanke off “China Focus” list and the following day, Goldman Sachs scrapped Vanke off the “Conviction” list. An Bang, which had strongly affirmed its support for Wang Shi raised its stake further to 7.01\% on December 17 despite these warning signs. Fitch confirmed that Vanke’s ratings were immediately unaffected by the tussle and that it will monitor the developments closely. How prepared was Vanke to oust an unwavering adversary without actual ownership of the company? Vanke halted trade on December 18 at 13:00, citing asset restructuring and acquisition after the stock price had surged by 62\% in December alone. \\
\subsection{Vanke’s defense mechanism}
When Vanke suspended trading of its shares, its largest shareholders were Baoneng with 24.4\% stake, followed by China Resources, which had 15.3\%, An Bang was third at 7\%, Guosen Jinping controlled4.7\% while China Galaxy Securities owned 3.5\%, China Securities Finance Corp had 3.4\% and Citic Securities owned 3.1\% (see Figure 3). Before the takeover, most of Vanke’s fragmented share structure made it easy for a fierce conglomerate like Baoneng to make a bulk purchase of shares on the market. After halting trading, investors speculated that Vanke could use a poison pill strategy to dilute Baoneng’s stake. The strategy would have Vanke issue new shares to a private investor at a discount, thereby diluting not only Baoneng’s but also other major shareholders’ stake. Moreover, a buyback of shares from individual investors was a possible alternative, only executable if Vanke has the resources to purchase the overvalued shares.  Earlier in July 2015 when the hostile takeover began, Vanke set out 10 billion yuan to buy back shares but due to the surge in price, they could only repurchase 160 million worth of shares, accounting for 0.113\%. Other sources contemplated that China Resources would continue to increase controlling interest in Vanke to regain control as the largest shareholder.\\
\end{multicols}
\begin{center}
\begin{figure}[h]
\caption{Ownership Structure of Vanke}
\includegraphics [scale=1]{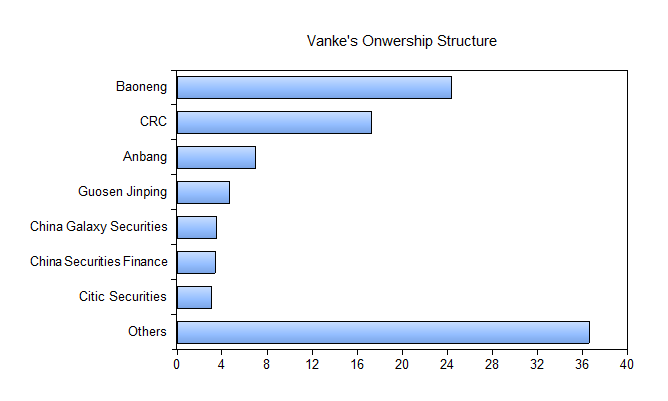} 
\end{figure}
\end{center}
\begin{multicols}{2}
\subsection{A White Knight}
Hong Kong ‘H’ shares resumed trading on January 6 and share price plummeted as investors feared that a reverse on the takeover attempt was looming. On March 14, Vanke announced its defense strategy. The developer had found a white knight, Shenzhen Metro Group Co. Ltd ("Shenzhen Metro"), to ease the pressure off Vanke. They signed an MOU with Shenzhen Metro agreeing to sell part of its assets to Vanke in exchange for shares and cash. The deal, estimated at 60 billion yuan was designed to have Vanke acquire Shenzhen Metro  for 45.6 billion yuan through offering 2.87 billion new shares at 15.88 yuan a piece in exchange for 20.65\% stake in Vanke. This seemingly amicable arrangement would dilute Baoneng’s stake to 19\%. When China Resources learnt that their stake was at risk of dilution, they disapproved the deal and blamed Vanke for not consulting them. In board meeting to decide the striking Shenzhen Metro deal, seven voted in favour of the deal and three board members from China Resources voted against it while one board member abstained. Vanke made a triumphant announcement of victory to proceed with the deal. China Resources grudgingly questioned the legality of the board’s resolution in that the rules required 2/3 majority for the deal to pass, and the 7/11 vote was insufficient. Though Jushenghua and Foresea Life were not invited to the board meeting, they unequivocally expressed their rejection of the proposal. After the vote, Baoneng criticizing Vanke's executives for mismanagement of the company and immediately called out to oust Wang Shi and all his board members. The decision to oust Vanke caused H shares to slip 3.8\% even though rating agencies had warned against it. Vendors began to renegotiate contract terms as more and more traders dumped the shares. The decision by China Resources and Baoneng to block Shenzhen Metro deal raised suspicion that the two were acting in concert. On June 27, CSRC asked whether Baoneng and China Resources were acting in concert. While China Resources had rejected the Shenzhen Metro deal, they opposed the proposal by Baoneng to oust Vanke’s management. They further clarified that China Resources did not reach an agreement with Baoneng to vote in concert. \\
\subsection{Vanke’s performance}
During mergers and acquisitions, company culture can change dramatically and employees may leave the company or lose their jobs. One of Vanke's Non-Executive Directors, tendered his resignation on December 22, 2016. Vanke denied allegations that the resignation was prompted by mismanagement of the takeover feud. On January 1, Vanke posted total sales of 262.7 billion yuan, making it the number one developer in China by sales revenue. This confirmed that Vanke’s ongoing battle for control did not eat into the company’s revenue and growth, (see Figure 1). In February, Vanke beat 17 developers to win a residential site in Hong Kong Government tender. At the end of the same month, the U.S unit of China Vanke joined with Slate Property Group and Adam America Real Estate to purchase 45 Rivington Street for U.S\$116 million. Back in mainland China, the developer began to incorporate a new business model dubbed “Rail + Property” by entering into strategic partnership with Shenzhen Metro Group and acquire its property division. As parties to the tussle fought tooth and nai to gain control, Vanke bought 96.55\% interest in property firms held by Blackstone, a U.S financial services company, at a cost of 3.89 billion yuan. \\
On July 22, a day after sealing the deal with Blackstone, Vanke announced to pay 2015 cash dividends of 7.20 yuan pretax, for every 10 shares held as of July 28, 2015. The 6-month January to July contract sales were 217.5 billion yuan, which was relatively higher, compared to the previous year. Even though the second quarter financial statements for Vanke showed the company’s performance had improved year on year, the third quarter results released in October showed that Evergrande had surpassed Vanke in total sales in the month of August. Vanke’s total sales grew by 49\% and the developer criticised Baoneng for reducing confidence among partners and customers. This led to project cancellations and tighter credit. The long run effects would cloud Vanke's performance and pose risk among shareholders. Investors rallied behind the feud, driving stock price high while paying superficial attention to operational and financial results.\\

\section{Battle for Control 2: The more the merrier}
On July 1, Vanke’s listed shares in Shenzhen resumed trading for the first time in more than six months, (see Figure 4). The anticipated restructuring process was widely viewed by investors as an opportunity grasped by Vanke to reverse the takeover. Out of fear, investors started ditching the A shares and they plunged 10\% - triggering a halt, to 21.99 yuan from 24.43 yuan which was the closing price on the day of suspension. On the other hand, H shares jumped significantly up by 8.4\% to HK\$16.48 on the same day. As shares resumed trading and price fell, Baoneng remained unfazed and seized the opportunity to increase stake to 24.972\% at a cost of 1.5 billion yuan. The regulations in mainland China does not allow shares to fluctuate by more than 10\% in a day, and in the event that daily volatility reaches 10\%, a halt will be triggered to suspend trading for that day. Another controversial rule on China’s equity markets is the T+1 which only allows investors to sell stocks at least after one day of holding it. In other words, investors are forbidden from selling stocks that they have bought on the same day. The T+1 comes after the CSRC abandoned the T+0 strategy which was implemented in 1995 but later abandoned in a bid to control market volatility and risk. The ongoing debate on this issue is whether T+1 reduces or exacerbates market volatility and market manipulation as well as protect the interests of investors.\\
\end{multicols}
\begin{center}
\begin{figure}[h]
\caption{Performance of Vanke's shares before and after suspension}
\includegraphics [scale=1]{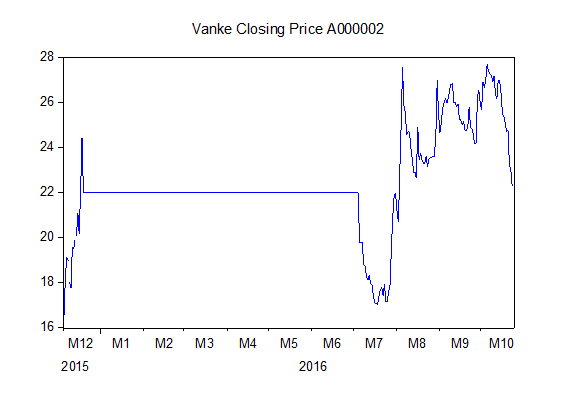} 
\end{figure}
\end{center}
\begin{multicols}{2}
\subsection{China Evergrande joins the takeover battle}
On August 4, Wang Shi was rattled by the news that Evergrande Real Estate Group Ltd ("Evergrande Group"), a Fortune500 developer started in Guangzhou in 1996 had bought 4.68\% stake in his company, making it the fourth largest shareholder in Vanke. The purchase of 516.9 million A shares at a total cost of 9.1 billion yuan had Vanke stock reaching the 10\% volatility peak, closing at 19.67 yuan in Shenzhen, while H shares surged 3\% and closed at HK\$18.60. Rumors that Vanke leaked information about Evergrande’s share purchase before official announcement attracted the attention of China Securities Regulatory Commission (CSRC), which issued a statement for Vanke to provide an explanation; however, Vanke denied the allegations. In 2015, Evergrande Group's total sales reached 201.34 billion, making it China's second largest developer by sales. Evergrande group cited strong results as a reason for the share purchase. After a whole year of the hostile takeover, a consortium of companies were still willing to join the battle for control. As more developers increased their participation, China stocks rose in the month of August in other sectors such as coal. Within a week, Evergrande Group increased its stake to 6.82\% at a cost of 14.57 billion yuan, overtaking An Bang and becoming the third largest shareholder. Evergrande Group announced that its share purchase in Vanke was made out of its cash reserves despite bearing a 600\% debt level that could have negative impact on Vanke’s ratings. On August 24, S\&P revised Vanke’s outlook from stable to negative though prevailing credit rating remained unchanged. Moody followed suit and slashed Vanke to negative due to uncertainty in management and tightened credit conditions. \\
During the first week of November, Evergrande Group increased in Vanke stake to 8.3\%. On November 11, CSRC suspended Evergrande Group's trading accounts due to “abnormal trading behavior”, a tendency to snap up publicly traded stocks of other companies on secondary market. Furthemore, Shenzhen Stock Exchange issued a warning to Evergrande Group to stop using its insurance unit as a slush for trading assets. China Insurance Regulation Commission ("CIRC") followed by expressing its unsupportive stance to Evergrande Life Insurance and further summoned it to refrain from the abnormal behavior. In a filing to Hong Kong Stock Exchange on November 18, Vanke reported that Evergrande had acquired 1.04 billion A shares, boosting their stake to 9.452\% at a total cost of 22.26 billion yuan. On the same day, Vanke shares hit a record high of 28.9 yuan per piece. New World Development, a Hong Kong based developer was also secretly snapping Vanke’s A shares on the open market and their position remained unknown. The feud got intense, billions of yuan kept pouring and the lack of transparency exacerbated the battle. The regulations require that when a company holds 5\% interest in another entity, it may request a seat on the board of directors.\\
\subsection{How Baoneng financed the acquisition}
After purchasing about 25.4\% of Vanke at an estimated cost of 43 billion yuan, Baoneng was by far Vanke's largest shareholder. However, Wang Shi was determined to fight for his company. On July 17, Vanke asked CSRC to investigate Baoneng on allegations of misconduct. Baoneng, through its subsidiary Jushenghua, had set up nine Asset Management Plans ("AMP") (see Figure 3), a type of shadow banking arrangement used to finance stock purchases and takeovers. JP Morgan \& Chase estimated that 26 billion of the 43 billion yuan was lend by six banks through AMP, and  that AMP funds in China had already reached 32 trillion yuan. The banks listed by Vanke to have participated in setting up the AMPs included Ping An Bank, Guangfa Bank, China Construction Bank and China Mingsheng Bank. In addition to that, Vanke stated that Baoneng did not qualify to be registered as a shareholder because they violated regulations. 
The courage by Wang Shi to continue battling the takeover sent the price falling and between July 4th and July 19th, Vanke shares had collapsed by 30\%, threatening margin calls and more chaos as banks began to seek an exit out of the funding arrangements. Baoneng was at the verge of a forced liquidation.
In response to the investigation, Shenzhen Stock Exchange found that Vanke and its shareholder Jushenghua had violated disclosure regulation, and further issued a letter requesting the companies to comply. The CSRC summoned directors of both companies to meet with the regulatory body and discuss the matter. The commission expressed that Vanke and Jushenghua had disregarded the stability of capital markets, sustainable development of the firm and interest of small shareholders. \\
The growing trend that conglomerates were using their insurance arms as vehicles of financing takeovers and acquisitions prompted China Insurance Regulatory Commission (CIRC) to issue a stern warning against the behavior. A collaborated effort by CSRC, CIRC and China Banking Regulatory Commission ("CBRC") to investigate Vanke’s takeover battle found no irregularities in Baoneng’s fundraising mechanism. The investigation gave birth to more regulations which tightened corporate fundraising. On July 27, a new regulation was passed which stipulated that shareowners who own at least 5\% of a listed firm could not use money netted through the issuance of Wealth Management Products ("WMPs"), AMPs or capital raised from third party fundraising platforms to buy additional shares in the company. \\

\section{A closer look at major players}
The Vanke takeover battle did not only test the equity markets in China, but also revealed the extent to which government interferes in corporate takeovers. Companies that are well connected to powerful officials often rely on government support when market dynamics are deemed disruptive. On one hand, before he founded Vanke, Wang Shi was once a railway official married to the daughter of the then deputy Communist Party boss of Guangdong province, Wang Ning. When Wang Shi resigned from the position of general manager in 1999, he appointed Yu Liang who left a major state owned enterprise to be his successor. During the takeover battle, Vanke had the backing of An Bang Insurance Group, which was the third largest shareholder of Vanke before Evergrande joined the tussle. The founder of An Bang, Wu Xiaohui is married to the granddaughter of Deng Xiaoping – the communist leader credited with pioneering China’s opening up policy in 1978. It can be argued that the success of major companies depends on the relationship that exist between the founders and officials in positions of power.\\
On the other hand, Baoneng Group was set up by two brothers, Yao Zhenhua an elder brother of Yao Jianhui, who both control Foresea Life and Jushenghua. Baoneng Group, established in 1992 first traded as a retailing business but further developed into a massive conglomerate through acquisitions and it now owns over 40 shopping malls. Its core businesses include real estate development, modern logistic industry, cultural tourism and financial industry. China Resources, a multi-business holding state-owned enterprise established in Hong Kong in 1938 as Liow \& Company but later renamed China Resources Company in 1948. The company was the largest shareholder of Vanke for the past 15 years before the takeover battle with 14.89\% stake. As the battle continued, a non-executive director of Vanke openly criticized China Resources for lending cash to Baoneng to finance its acquisition in Vanke. Baoneng was reported to have pledged 2.02 billion shares in Jushenghua to a division of China Resources in July 2015. In response, China Resources stated that the share pledge was to offset an outstanding payment from Baoneng on a previous real estate project that was jointly developed by Baoneng and its subsidy, China Resources Land in 2015. Whether China Resources and Baoneng were acting in concert from the first day of the tussle remains unknown to the public.\\

\section{A Comparison with other takeovers}
In China, previous acquisitions have not been as confrontational as the Vanke’s case. Given that the financial markets in China are only 30 years old, the business environment does not support much activism that can lead to hostile takeover as compared to mature markets. Most corporate takeovers in China are settled peacefully through negotiations with the help of local and provincial authorities in order to minimize costs and losses and ensure stability of the markets. Further transformation in China’s financial markets is likely to see increased activism and corporate takeovers. In fact, Vanke was not the only company to undergo corporate takeover in 2015. Tianrui International Holding Co. Ltd increased its stake in the major cement maker China Shanshui Cement Group Ltd to 28\% in the open market without declaring this to its target. The tussle went on for several months after which China Shanshui defaulted and Tianrui International successfully removed China Shanshui’s management board. In 2004, Japan’s Sumitomo Mitsui Financial Group failed to acquire UFJ Holdings when Mitsubishi UFJ Financial Group crashed the takeover efforts. However, outside China, Chinese firms have become popular for huge acquisitions and investments in real estate, football clubs and high tech firms. In U.S equity markets, corporate takeovers began as early as 1893 characterized by horizontal mergers where firms in manufacturing and mining were combining to eliminate competition. Early examples of U.S acquisitions include Standard Oil Company of New Jersey founded in 1870, which made a series of acquisitions and finally became known as the New Jersey Holding Company; followed by United States Steel Corporation in 1901. Another great wave of hostile mergers in U.S happened after the dotcom boom in early 1990s and they were characterized by Leveraged Buy-outs. \\

\section{Farewell to Arms}
After fighting relentlessly for a year and a half, Wang Shi managed to bring back Shenzhen Metro, the same company whose 60 billion yuan deal was blocked by China Resources Company in a vote in June 2016. On 13 January 2017, Shenzhen Metro announced that it would buy the entire 15.3\% stake of Vanke’s second largest shareholder, China Resources for 37.17 billion yuan, thus, an average of 22 yuan per share. Part of the agreement also meant that Shenzhen Metro would take over the three board seats in Vanke, which were previously held by China Resources. Apart from that, in June 2017, Shenzhen Metro announced plans for a major acquisition of Vanke, which led to the suspension of Vanke shares for a week from June 7 to 11. The restructuring plan saw Shenzhen Metro acquiring 1.55 billion A-shares at a cost of 18.80 yuan per share (a total of 29.2 billion yuan) from Evergande’s subsidiary Hengda Real Estate Group that controlled 14.07\% of Vanke. The restructuring made Shenzhen Metro the largest shareholder of Vanke garnering 29.37\% control in total.This deal marked the end of the hostile takeover for Vanke. On June 21, 2017, Wang Shi announced his resignation as Chairman of Vanke. The company’s president Yu Liang took over as the Chairman of the real estate conglomerate. On the same day, Shenzhen Metro, Vanke’s largest shareholder proposed nominees for board of directors of whom three were original Vanke directors. Shenzhen Metro did not nominate any directors from second or third largest shareholders.  An investigation by CIRC revealed that Yao Zhenhua, Chairman of Baoneng  had used funds from his insurance company to undertake a hostile takeover. On January 27, Yao Zhenhua was stripped off his leadership position in Foresea Life Insurance and effectively barred from the insurance industry for a period of 10 years. CIRC also banned Evergrande Life, the insurance arm of Evergrande Group, from investing in stocks.\\
On June 30, the new board was sworn in, headed by Yu Liang as the new chairman of the board. Although Baoneng Group did not show up at the swearing in meeting, Vanke announced that Baoneng Group had expressed support for the new board. The new board agreed that, Vanke, in partnership with Shenzhen Metro would continue to focus on providing integrated urban services, expanding realty business partnership to property management. The board also put up a plan to build 32 lines in total in the next 10 to 15 years as Shenzhen Metro and Vanke continue their expansion into second and third tier cities. \\

\section{Conclusion}
The hostile takeover became a yardstick for development in China’s financial markets. In the context of China, the markets are young but with improved regulation and transformation, other foreign firms will become prevalent players in the market. To a greater extent, hostile takeovers in China are generally not welcome. Vanke became an exceptional case as one of the testing ground for hostile takeovers on Chinese soil. The takeover attracted major players in the financial markets including banks, hedge funds, public and private companies. It is interesting to note while the government had to leave the battle to market participants, it also had a duty to guarantee a level playing field for all players. Whether the government carried out its responsibility without bias is subject to debate, given that Vanke’s ‘white night’ was a state-owned conglomerate. The Vanke takeover battle could mark the genesis of more hostile takeovers in China; a phenomenon that United States has experienced since 1890s.

\end{multicols}

\end{document}